\documentclass[onecolumn,preprint,a4paper,nofootinbib]{revtex4}
\usepackage[mathscr]{euscript}
\usepackage{ulem}
\usepackage{caption}
\usepackage{amsmath}
\usepackage{multirow}
\usepackage{booktabs,tabularx}
\usepackage{threeparttable}
\usepackage{floatrow}
\usepackage[font=small,labelfont=bf,tableposition=top]{caption}
\usepackage{booktabs}
\usepackage{threeparttable}
\usepackage{graphicx}
\usepackage{subfigure}
\usepackage{epstopdf}
\usepackage[colorlinks=true,linkcolor=red]{hyperref}
\usepackage{bm}
\usepackage[figuresright]{rotating}
\allowdisplaybreaks[3]
\newcommand{\bea}{\begin{eqnarray}}
\newcommand{\eea}{\end{eqnarray}}
\newcommand{\beq}{\begin{equation}}
\newcommand{\eeq}{\end{equation}}

\def\/{\over}

\begin{document}
\title{Gravitational Casimir-Polder interaction in a thermal bath}

\author{Shijing Cheng\footnote{Corresponding author: chengshijing@sxnu.edu.cn}, Tianxin Wang, and Zili Zhang}
\affiliation{School of Physics and Electronic Engineering, Shanxi Normal University, Taiyuan 030031, China}

\begin{abstract}

We have investigated, by separating the contributions from thermal fluctuations (tf) and the radiation reaction (rr), the gravitational Casimir-Polder interaction between a gravitationally polarizable two-level object and an infinite gravitational Dirichlet boundary in a thermal bath at a temperature $T$.
The results indicate that the rr-contribution to the interaction potential is independent of the temperature, whereas the tf-contribution is generally governed by a nontrivial interplay between the thermal corrections and the polarization effect. 
Here, the object-to-boundary distance, the characteristic transition wavelength of the object and the thermal wavelength of gravitons are denoted by $L$, $\lambda$ and $\beta$, respectively.
In contrast to the vacuum case, where the interaction potential scales as $L^{-5}$ for $L\ll\lambda$ and $L^{-6}$ for $L\gg\lambda$, corresponding to an always repulsive force, qualitatively new behaviors emerge at high temperatures.
Particularly, when $\sqrt[4]{\beta\lambda^3}\ll L\ll\lambda$ and the object is polarizable within the plane perpendicular to the boundary, a novel scaling of $TL^{-1}$ arises; when $\sqrt{\beta\lambda}\ll L\ll\lambda$ and the object is polarizable along the vertical-to-boundary axis, the interaction force becomes surprisingly attractive.
At extremely high temperatures and large distances, i.e. when $\beta\ll \lambda\ll L$, the potential oscillates with the distance $L$ and thus an attractive or repulsive and even vanishing force can be resulted, depending on the exact values of $L$.
Our work demonstrates that thermal gravitons can act as an active control mechanism for quantum gravitational interactions, and temperature, polarization configuration, and object-to-boundary distance jointly determine the magnitude, scaling law, and even the attractive or repulsive nature of the interaction force.

\end{abstract}

\maketitle

\section{Introduction}

The Casimir and Casimir-Polder (CP) interactions, arising from vacuum fluctuations of electromagnetic fields, are representative manifestations of quantum electrodynamics in the presence of boundaries (see Refs.~\cite{Bordag01,Klimchitskaya09,Bordag09} for extensive reviews).
The CP interaction was originally calculated in the situation of an electrically neutral ground-state atom and a perfectly conducting plate~\cite{Casimir48}.
When the atom-to-plate distance $L$ is much smaller than the characteristic transition wavelength of the atom, the CP interaction reduces to the van der Waals potential with a characteristic behavior of $L^{-3}$.
However, at the larger distances, such interaction decays more rapidly and scales as $L^{-4}$ as a result of the retardation effect~\cite{Casimir48}.
It has later been demonstrated within the framework of Lifshitz theory that, in addition to the zero-point fluctuations, the thermal fluctuations can also generate dominant modifications~\cite{Lifshitz56,Dzyaloshinskii61}.
Specifically, the CP interaction is governed by vacuum fluctuations only at short distances and low temperature; however, when the atom-to-surface distance exceeds the wavelength of thermal photons, it acquires a classical thermal contribution proportional to $TL^{-3}$, where $T$ denotes the environmental temperature.
Subsequent investigations have further demonstrated the CP interactions in various situations, such as with different boundary conditions~\cite{Zhou95,Eberlein07,Wu2000,Milton11}, or the atom in noninertial motion in vacuum~\cite{Rizzuto07,Zhu10,Rizzuto09} or static in the environment in or out of thermal equilibrium~\cite{Zhu09,Antezza05,Antezza06,Buhmann08,Obrecht07,She10}.

In parallel, when gravity is quantized, one may also expect analogous effects arising from quantum fluctuations of gravitational fields.
Although a complete theory of quantum gravity remains unavailable at present, the low-energy effects can be consistently explored within the framework of linearized quantum gravity or effective field theory.
In recent years, increasing attentions have been devoted to the gravitational-fluctuation-induced effects, such as the flight time fluctuations of a probe light signal from its source to a detector ~\cite{Ford96,Yu99,Yu09}, the entanglement generation between two initially independent systems~\cite{Cheng18}, and the spontaneous excitation of an accelerated atom~\cite{Cheng19}.
In analogy with the electromagnetic CP interaction induced by electric dipoles, another expected effect is the quantum gravitational interactions among gravitationally polarizable objects and gravitational boundaries which originate from mass quadrupole moments induced by gravitational vacuum fluctuations~\cite{Quach15,Ford16,Pinto16,Wu16,Hu25,Hu17,Wen25,Cheng26}.
Existing studies have shown that the gravitational CP interaction between a ground-state object and a gravitational Dirichlet boundary exhibits characteristic distance-dependent behaviors at the short and large distances, scaling as $L^{-5}$ and $L^{-6}$ respectively~\cite{Hu17}.
Here it should be emphasized that ordinary materials generally possess extremely weak reflectivity for gravitational waves~\cite{Smolin85}, whereas certain quantum matter, such as superconducting films, can serve as effective gravitational mirrors and thus mimic gravitational Dirichlet boundary condition~\cite{Minter10,Chiao17}.
More recently, the gravitational CP interactions in the cases of an excited object or in presence of a gravitational Neumann boundary have also been investigated, and significantly the attractive or repulsive property of the interaction force could be changed~\cite{Wen25,Cheng26}.

The above-mentioned studies on the gravitational CP interaction are restricted to vacuum settings, and then it is natural to ask how gravitational thermal fluctuations affect this interaction.
In this paper, we intend to study the gravitational CP interaction in a thermal bath at the constant temperature.
Particular attention is devoted to the influence of temperature on the interaction potential in different distance regions with various polarizations, and also to the respective roles played by gravitational thermal fluctuations and radiation reaction of the object.
For this purpose, we resort to the formalism proposed by Dalibard, Dupont-Roc, and Cohen-Tannoudji (DDC)~\cite{Dalibard82,Dalibard84}, which provides a transparent separation between the contributions originating from field fluctuations and radiation reaction of the system itself.
In recent years, the DDC method has proved particularly powerful in the investigations of spontaneous emission~\cite{Audretsch94,Zhu07,Zhu06,Jin14}, Lamb shifts~\cite{Audretsch95,Holzmann95,Tomazelli03}, the fluctuation-induced interactions~\cite{Rizzuto16,Cheng2022,Zhou1897,Zhou21,Noto14,Menezes17}, and the collective transitions of two entangled atoms~\cite{Menezes15,Menezes16,Zhou20}.
This paper is divided into five sections.
In section II we give the general equations for the contributions of thermal fluctuations and radiation reaction to the gravitational CP interaction, then in section III concretely calculate the results in the case of the ground-state object placed near the gravitational Dirichlet boundary in a thermal bath.
In section IV, we discuss the detailed behaviors of the interaction potential in the low- and high-temperature regimes respectively.
The summary of this paper is presented in section V.

\section{Basic formalism}

We consider a non point-like two-level object interacting with a bath of fluctuating gravitational fields in the presence of an infinite Dirichlet boundary plane.
For such system, the total Hamiltonian consists of three parts: the Hamiltonian of the object $H_S$, the Hamiltonian of the gravitational fields $H_F$, and the interaction Hamiltonian $H_I$.
Explicitly,
\bea
H_{S}(\tau)&=&\sum_n\omega_n \sigma_{nn}(\tau)\;,\label{HS}\\
H_{F}(\tau)&=&\sum_{\lambda}\int d^3\textit{\textbf{k}}\ \omega_{\textit{\textbf{k}}}a^{\dag}_{\textit{\textbf{k}},\lambda}(t(\tau))a_{\textit{\textbf{k}},\lambda}(t(\tau))\frac{dt}{d\tau}\;,\label{HF}\\
H_{I}(\tau)&=&-\frac{1}{2}Q_{ij}(\tau)E_{ij}(\mathrm{x}(\tau))\;.
\eea
Here, $\tau$ and $t$ are proper time and coordinate time respectively, $\sigma_{nn}=|n\rangle\langle n|$ with $|n\rangle$ denoting the ground state $|g\rangle$ or the excited state $|e\rangle$ with energy $\omega_n$, $a^{\dag}_{\textit{\textbf{k}},\lambda}$ and $a_{\textit{\textbf{k}},\lambda}$ are respectively the creation and annihilation operators with $\lambda$ the polarization index and $\textit{\textbf{k}}$ the wave vector, $Q_{ij}$ is the mass quadrupole moment operator of the object, and $E_{ij}$ denotes the gravito-electric tensor defined as $E_{ij}=-\nabla_i\nabla_j\phi$ with $\phi$ the gravitational potential.

As a result of the boundary-induced modifications on quantum gravitational fluctuations, the energy shifts of the object become dependent of the distance between the object and the boundary.
This distance dependence then gives rise to the gravitational Casimir-Polder interaction, and the energy shifts of the object dependent of the distance only exactly correspond to the interaction potential.
In Ref.~\cite{Cheng26}, the general contributions of vacuum gravitational fluctuations and radiation reaction of the ground-state object to the interaction are obtained.
Following the similar procedures, and taking the expectation values of the change rate of the energy for the object [i.e. Eqs.~(13) and (14) in Ref.~\cite{Cheng26}] on the thermal state $|\beta\rangle$ instead of $|0\rangle$, one can easily obtain the contributions of thermal fluctuations (tf) and radiation reaction of the object (rr) to the gravitational CP potential as
\bea\label{tf1}
(\delta E)_{tf}=-\frac{i}{4}\int_{\tau_0}^{\tau}d\tau'C^F_{ijkl}\left(\mathrm{x}(\tau),\mathrm{x}(\tau')\right)\chi^A_{ijkl}(\tau,\tau')\;,
\eea
and
\bea\label{rr1}
(\delta E)_{rr}=-\frac{i}{4}\int_{\tau_0}^{\tau}d\tau'\chi^F_{ijkl}\left(\mathrm{x}(\tau),\mathrm{x}(\tau')\right)C^A_{ijkl}(\tau,\tau')\;,
\eea
respectively.
Here, $C_{ijkl}^A$ and $\chi_{ijkl}^A$ denote the symmetric and antisymmetric statistical functions of the object prepared in the state $|b\rangle$, defined by
\bea
C_{ijkl}^A(\tau,\tau')&=&\frac{1}{2}\langle b|\left\{Q_{ij}^f(\tau),Q_{kl}^f(\tau')\right\}|b\rangle,\label{atomC}\\
\chi_{ijkl}^A(\tau,\tau')&=&\frac{1}{2}\langle b|\left[Q_{ij}^f(\tau),Q_{kl}^f(\tau')\right]|b\rangle\;,\label{atomX}
\eea
with the superscript ``$f$'' denoting free operator, and they only depend on the object itself.
Similarly, the gravitational-field symmetric and antisymmetric correlation functions $C_{ijkl}^F$ and $\chi_{ijkl}^F$\;, are defined by
\bea
C_{ijkl}^F(\mathrm{x}(\tau),\mathrm{x}(\tau'))&=&\frac{1}{2}\langle \beta|\left\{E^f_{ij}(\mathrm{x}(\tau)),E^f_{kl}(\mathrm{x}(\tau'))\right\}|\beta\rangle\;,\label{fieldC}\\
\chi_{ijkl}^F(\mathrm{x}(\tau),\mathrm{x}(\tau'))&=&\frac{1}{2}\langle \beta|\left[E^f_{ij}(\mathrm{x}(\tau)),E^f_{kl}(\mathrm{x}(\tau'))\right]|\beta\rangle\;,\label{fieldX}
\eea
and here only the distance-dependent parts lead to the gravitational CP force.

\section{The gravitational Casimir-Polder interaction potential in a thermal bath}

We assume that the object is static in a thermal bath at the temperature $T$, with its coordinates given by
\bea\label{tra}
t(\tau)=\tau,\ x(\tau)=y(\tau)=0,\ z(\tau)=L\;,
\eea
and an infinite Dirichlet boundary coincides with the $xoy$ plane, thus the distance between the object and the boundary is $L$.

According to the generic expressions~(\ref{tf1}) and (\ref{rr1}), the evaluation of the gravitational CP interaction potential requires the determination of four statistical functions $C_{ijkl}^F$\;, $\chi_{ijkl}^F$\;, $C_{ijkl}^A$ and $\chi_{ijkl}^A$\;.

Under the weak-field approximation, the spacetime metric can be expressed as the sum of the flat spacetime metric and a linearized perturbation $h_{\mu\nu}$.
In the presence of the infinite Dirichlet boundary, the quantized metric perturbation $h_{ij}$ in the transverse-traceless gauge takes the following form
\bea\label{hij}
h_{ij}(\textit{\textbf{x}},t)=\int d^3\textit{\textbf{k}}\sum_{\lambda}\biggl\{\frac{a_{\textit{\textbf{k}},\lambda}}{2\sqrt{(2\pi)^3\omega}}\bigl[e_{ij}(\textit{\textbf{k}},\lambda)e^{i(\textit{\textbf{k}}\cdot\textit{\textbf{x}}-\omega t)}-e_{ij}(\textit{\textbf{k}}^-,\lambda)e^{i(\textit{\textbf{k}}^-\cdot\textit{\textbf{x}}-\omega t)}\bigl]+\mathrm{H.c.}\biggl\}\;,
\eea
where $i,j$ denote spatial indices, $e_{ij}(\textit{\textbf{k}},\lambda)$ is the polarization tensor, $\textit{\textbf{k}}=\{k_x,k_y,k_z\}$, $\textit{\textbf{k}}^{-}=\{k_x,k_y,-k_z\}$, $\omega=|\textit{\textbf{k}}|=|\textit{\textbf{k}}^-|$, and $\mathrm{H.c.}$ denotes the Hermitian conjugate.
In general relativity, the gravito-electric tensor $E_{ij}$ can be defined in terms of the Weyl tensor $C_{i0j0}$, and can thus be expressed as
\bea\label{E}
E_{ij}=\frac{1}{2}\ddot{h}_{ij}\;,
\eea
where a dot means the derivative with respect to time.
Using Eqs.~(\ref{hij}) and (\ref{E}), the two-point correlation function of the gravitational fields in the thermal state, which only depend on the object-to-boundary distance $L$, can be derived as
\bea\label{Eijkl}
&&\langle \beta|E^f_{ij}(\mathrm{x})E^f_{kl}(\mathrm{x}')|\beta\rangle =-\frac{1}{128\pi^3}\int d^3\textit{\textbf{k}} \nonumber\\
&&\times\biggl\{\frac{\omega^3e^{\beta\omega}}{e^{\beta\omega}-1} e^{-i\omega(\tau-\tau')}\sum_{\lambda}\biggl[e_{ij}(\textit{\textbf{k}},\lambda)e_{kl}(\textit{\textbf{k}}^-,\lambda)e^{ i\textit{\textbf{k}}\cdot\textit{\textbf{r}}}+e_{ij}(\textit{\textbf{k}}^-,\lambda)e_{kl}(\textit{\textbf{k}},\lambda)e^{ i\textit{\textbf{k}}^-\cdot\textit{\textbf{r}}}\biggl]\nonumber\\
&&\quad \ +\frac{\omega^3}{e^{\beta\omega}-1} e^{i\omega(\tau-\tau')}\sum_{\lambda}\biggl[e_{ij}(\textit{\textbf{k}},\lambda)e_{kl}(\textit{\textbf{k}}^-,\lambda)e^{ -i\textit{\textbf{k}}\cdot\textit{\textbf{r}}}+e_{ij}(\textit{\textbf{k}}^-,\lambda)e_{kl}(\textit{\textbf{k}},\lambda)e^{ -i\textit{\textbf{k}}^-\cdot\textit{\textbf{r}}}\biggl]\biggl\}\;,
\eea
where $\textit{\textbf{r}}=\{0,0,2L\}$.
In obtaining Eq.~(\ref{Eijkl}), we have used the following two equations
\bea
\langle\beta|a_{\textit{\textbf{k}},\lambda}a^{\dag}_{\textit{\textbf{k}},\lambda}|\beta\rangle=1+\frac{1}{e^{\beta\omega}-1}\;,\qquad\qquad
\langle\beta|a^{\dag}_{\textit{\textbf{k}},\lambda}a_{\textit{\textbf{k}},\lambda}|\beta\rangle=\frac{1}{e^{\beta\omega}-1}\;,
\eea
where $\beta=1/T$ denotes the thermal wavelength.
The summation over polarization tensors in Eq.~(\ref{Eijkl}), according to Ref.~\cite{Yu99}, is given by
\bea
\sum_{\lambda}e_{ij}(\textit{\textbf{k}},\lambda)e_{kl}(\textit{\textbf{k}}^-,\lambda)&=&\delta^-_{ik}\delta^-_{jl}+\delta^-_{il}\delta^-_{jk}-\delta_{ij}
\delta_{kl}
-\delta^-_{ik}\hat{k}_j\hat{k}^-_l-\delta^-_{jl}\hat{k}_i\hat{k}_k^--\delta^-_{il}\hat{k}_j\hat{k}_k^-\nonumber\\
&&-\delta^-_{jk}\hat{k}_i\hat{k}_l^-+\delta_{ij}\hat{k}^-_k\hat{k}_l^-+\delta_{kl}\hat{k}_i\hat{k}_j+\hat{k}_i\hat{k}_j\hat{k}^-_k\hat{k}_l^-\;,
\eea
where $\delta_{ij}$ is the Kronecker symbol, $\delta^-_{ij}=\delta_{ij}-2\delta_{zi}\delta_{zj}$, $\hat{k}_i=k_i/\omega$, $\hat{k}^-_i=k^-_i/\omega$.
By substituting $e^{i\textit{\textbf{k}}\cdot\textit{\textbf{r}}}$ with $e^{ 2i\omega L\cos{\theta}}$ and performing the integration in spherical coordinates,
\bea
\int d^3\textit{\textbf{k}}\rightarrow\int_0^{\infty}\omega^2d\omega\int_0^{\pi}\sin{\theta}d\theta\int_0^{2\pi}d\phi\;,
\eea
Eq.~(\ref{Eijkl}) can be further simplified.
Then, by employing the definitions in Eqs.~(\ref{fieldC})-(\ref{fieldX}), the correlation functions of the gravitational fields can be expressed as
\bea
C_{ijkl}^F(\mathrm{x},\mathrm{x}')&=&-\frac{1}{128\pi^2L^5}\int_0^{\infty}d\omega \mathcal{G}_{ijkl}(\omega L)\cos{( \omega\Delta \tau)}\left(1+\frac{2}{e^{\beta\omega}-1}\right)\;,\label{fieldc}\\
\chi_{ijkl}^F(\mathrm{x},\mathrm{x}')&=&\frac{i}{128\pi^2L^5}\int_0^{\infty}d\omega\mathcal{G}_{ijkl}(\omega L) \sin{(\omega\Delta \tau)}\;,\label{fieldx}
\eea
with $\Delta \tau=\tau-\tau'$.
Here, the function $\mathcal{G}_{ijkl}(\omega L)$ defined as
\bea
\mathcal{G}_{ijkl}(\omega L)\equiv f_{ijkl}(\omega L)\cos{(2\omega L)}+g_{ijkl}(\omega L)\sin{(2\omega L)}\;,
\eea
satisfies the symmetry relations listed below,
\bea
\left\{
  \begin{array}{ll}
\mathcal{G}_{xxxx}=\mathcal{G}_{yyyy}\;,\quad \mathcal{G}_{xxyy}=\mathcal{G}_{yyxx}\;,\\
\mathcal{G}_{xxzz}=\mathcal{G}_{yyzz}=\mathcal{G}_{zzxx}=\mathcal{G}_{zzyy}\;,\\
\mathcal{G}_{xyxy}=\mathcal{G}_{yxyx}=\mathcal{G}_{xyyx}=\mathcal{G}_{yxxy}\;,\\
\mathcal{G}_{xzxz}=\mathcal{G}_{zxzx}=\mathcal{G}_{xzzx}=\mathcal{G}_{zxxz}=\mathcal{G}_{yzyx}=\mathcal{G}_{zyzy}=\mathcal{G}_{yzzy}=\mathcal{G}_{zyyz}\;,
\end{array}
\right.
\eea
and the nonvanishing components of $f_{ijkl}(\omega L)$ and $g_{ijkl}(\omega L)$ are given by
\bea\label{fg}
\left\{
  \begin{array}{ll}
f_{xxxx}(s)=\frac{1}{2}s(8s^2-9),\\
f_{zzzz}(s)=-12s,\\
f_{xxyy}(s)=-\frac{1}{2}s(8s^2+3),\\
f_{xxzz}(s)=6s,\\
f_{xyxy}(s)=\frac{1}{2}s(8s^2-3),\\
f_{xzxz}(s)=2s(2s^2-3),
\end{array}
\right.\qquad
\left\{
  \begin{array}{ll}
g_{xxxx}(s)=\frac{1}{4}(16s^4-20s^2+9),\\
g_{zzzz}(s)=-2(4s^2-3),\\
g_{xxyy}(s)=-\frac{1}{4}(16s^4-4s^2-3),\\
g_{xxzz}(s)=4s^2-3,\\
g_{xyxy}(s)=\frac{1}{4}(16s^4-12s^2+3),\\
g_{xzxz}(s)=-3(2s^2-1).
\end{array}
\right.
\eea

We assume that the object is in its ground state.
According to Eqs.~(\ref{atomC})-(\ref{atomX}), the statistical functions of the object are calculated by
\bea
C_{ijkl}^A(\tau,\tau')&=&q_{ij}q^*_{kl}\cos{(\omega_{0}\Delta\tau)}\;,\label{atomc}\\
\chi_{ijkl}^A(\tau,\tau')&=&-i q_{ij}q^*_{kl}\sin{(\omega_{0}\Delta\tau)}\;,\label{atomx}
\eea
where $\omega_0$ is the energy gap and $q_{ij}\equiv\langle g|Q^f_{ij}(\tau_0)|e\rangle$.
Substitute Eqs.~(\ref{fieldc}), (\ref{fieldx}), (\ref{atomc}) and (\ref{atomx}) into the general expressions~(\ref{tf1}) and (\ref{rr1}), perform the integration over $\tau'$, and take the long-time limit $(\tau-\tau_0)\rightarrow\infty$.
Then the remaining frequency integrals can be evaluated by using contour integration and the residue theorem.
In this way, the contribution of thermal fluctuations to the gravitational CP interaction potential is obtained as
\bea\label{tf2}
(\delta E)_{tf}&=&\frac{|q_{kl}|^2}{256\pi^2 L^5}\int_0^{\infty}du \frac{\omega_0}{\omega^2_0-u^2}\frac{1}{e^{\beta u}-1}\bigl[f_{klkl}(uL)\cos{(2uL)}+g_{klkl}(uL)\sin{(2uL)}\bigl]\nonumber\\
&&+\frac{|q_{kl}|^2}{512\pi^2 L^5}\int_0^{\infty}du \frac{\omega_0}{\omega^2_0+u^2}e^{-2uL}\bigl[if_{klkl}(iuL)+g_{klkl}(iuL)\bigl]\nonumber\\
&&+\frac{|q_{kl}|^2}{1024\pi L^5}\bigl[f_{klkl}(\omega_0 L)\sin{(2\omega_0 L)}-g_{klkl}(\omega_0 L)\cos{(2\omega_0 L)}\bigl]\;,
\eea
with $|q_{kl}|^2\equiv q_{kl}q^*_{kl}$\;, and the contribution of the radiation reaction is given by
\bea\label{rr2}
(\delta E)_{rr}=-\frac{|q_{kl}|^2}{1024\pi L^5}\bigl[f_{klkl}(\omega_0 L)\sin{(2\omega_0 L)}-g_{klkl}(\omega_0 L)\cos{(2\omega_0 L)}\bigl]\;.
\eea
It can be seen from Eq.~(\ref{tf2}) that the tf-contribution naturally separates into two parts: the temperature-independent component (the last two terms on the right) which corresponds to the contribution induced by zero-point fluctuations, and the temperature-dependent component (the first term on the right) which arises from thermal effects.
In contrast, Eq.~(\ref{rr2}) shows that the rr-contribution is completely independent of temperature, and therefore coincides with the corresponding result in vacuum.

By summing the two contributions in Eqs.~(\ref{tf2}) and (\ref{rr2}), the total gravitational CP interaction potential in the thermal bath is obtained as
\bea\label{tot2}
(\delta E)_{tot}&=&\frac{|q_{kl}|^2}{256\pi^2 L^5}\int_0^{\infty}du \frac{\omega_0}{\omega^2_0-u^2}\frac{1}{e^{\beta u}-1}\bigl[f_{klkl}(uL)\cos{(2uL)}+g_{klkl}(uL)\sin{(2uL)}\bigl]\nonumber\\
&&+\frac{|q_{kl}|^2}{512\pi^2 L^5}\int_0^{\infty}du \frac{\omega_0}{\omega^2_0+u^2}e^{-2uL}\bigl[if_{klkl}(iuL)+g_{klkl}(iuL)\bigl]\;,
\eea
and obviously it reduces to the vacuum result in the zero-temperature limit ($\beta\rightarrow\infty$).
But generally, this interaction potential in the thermal bath can be interpreted as the vacuum result supplemented by thermal corrections.
In some special situations where the thermal revisions are dominating, novel behaviors of this interaction as compared to the vacuum case may appear.
The potential in Eq.~(\ref{tot2}) also depends explicitly on the orientations of the mass quadrupole moment of the object, as characterized by the components $q_{ij}$.
Specifically, the diagonal component $q_{xx}$, $q_{yy}$ or $q_{zz}$ corresponds to the mass distribution along the $x$, $y$ or $z$ axis respectively, and the off-diagonal components $q_{xy}$, $q_{xz}$ and $q_{yz}$ describe the mass distributions within the planes parallel to the $xoy$, $xoz$ and $yoz$ planes.

\section{Discussions on the gravitational Casimir-Polder interaction potential}

In this section, we present a detailed analysis of the tf-contribution to, the rr-contribution to as well as the total gravitational CP interaction potential in the thermal bath.
Since the general expressions given in Eqs.~(\ref{tf2})-(\ref{tot2}) are rather complicated and not particularly transparent, we perform analytical approximations in several limiting regimes in order to extract the underlying physical behaviors.
For convenience, we rewrite the interaction potentials in the form
\bea
(\delta E)_{h}&=&|q_{kl}|^2(\delta E)^{kl}_{h}\;,
\eea
where $h=tf,rr,tot$, and the components $(\delta E)^{kl}_{h}$ are given by
\bea
(\delta E)_{tf}^{kl}&=&\frac{1}{256\pi^2 L^5}\times\biggl\{\int_0^{\infty}du \frac{\omega_0\bigl[f_{klkl}(uL)\cos{(2uL)}+g_{klkl}(uL)\sin{(2uL)}\bigl]}{(\omega^2_0-u^2)(e^{\beta u}-1)}\nonumber\\
&&\quad\qquad\qquad+\frac{\pi\bigl[f_{klkl}(\omega_0 L)\sin{(2\omega_0 L)}-g_{klkl}(\omega_0 L)\cos{(2\omega_0 L)}\bigl]}{4}\nonumber\\
&&\quad\qquad\qquad+\int_0^{\infty}du \frac{\omega_0e^{-2uL}\bigl[if_{klkl}(iuL)+g_{klkl}(iuL)\bigl]}{2(\omega^2_0+u^2)}\biggl\}\;,\label{tf3}\\
(\delta E)_{rr}^{kl}&=&-\frac{1}{1024\pi L^5}\bigl[f_{klkl}(\omega_0 L)\sin{(2\omega_0 L)}-g_{klkl}(\omega_0 L)\cos{(2\omega_0 L)}\bigl]\;,\label{rr3}
\eea
and
\bea\label{tot3}
(\delta E)_{tot}^{kl}&=&\frac{1}{256\pi^2 L^5}\times\biggl\{\int_0^{\infty}du \frac{\omega_0\bigl[f_{klkl}(uL)\cos{(2uL)}+g_{klkl}(uL)\sin{(2uL)}\bigl]}{(\omega^2_0-u^2)(e^{\beta u}-1)}\nonumber\\
&&\quad\qquad\qquad+\int_0^{\infty}du \frac{\omega_0e^{-2uL}\bigl[if_{klkl}(iuL)+g_{klkl}(iuL)\bigl]}{2(\omega^2_0+u^2)}\biggl\}\;.
\eea
Recall that the boundary is located on the $xoy$ plane.
The diagonal components $(\delta E)_{h}^{xx/yy}$ and $(\delta E)_{h}^{zz}$ correspond to mass distributions along the axes parallel to or vertical to the boundary plane, respectively (referred to as parallel-axial or vertical-axial polarizations).
The off-diagonal components $(\delta E)_{h}^{xy}$ and $(\delta E)_{h}^{xz/yz}$ correspond to planar mass distributions parallel to or vertical to the boundary (referred to as parallel-planar or vertical-planar polarizations respectively).

To reveal the role of temperature, we consider two limiting cases: the low-temperature regime which means the temperature of the thermal bath $T$ much smaller than the energy gap of the object $\omega_0$, and the high-temperature regime in which $T\gg\omega_0$.
Introducing the characteristic transition wavelength of the object, $\lambda=2\pi/\omega_0$, these two limits correspond to $\beta\gg\lambda$ and $\beta\ll\lambda$, respectively.

\subsection{Low-temperature regime}

In the low-temperature limit ($\beta\gg\lambda$), three distinct distance regions can be identified depending on the relative magnitude of $L$, i.e., $L\ll\lambda\ll\beta$, $\lambda\ll L\ll\beta$ and $\lambda\ll\beta\ll L$.

\subsubsection{Short-distance region $L\ll\lambda\ll\beta$}

We begin with the short-distance region $L\ll\lambda\ll\beta$.
The corresponding results for the tf-contribution $(\delta E)_{tf}^{kl}$\;, the rr-contribution $(\delta E)_{rr}^{kl}$ and the total gravitational CP potential $(\delta E)_{tot}^{kl}$\;, obtained from the general expressions in Eqs.~(\ref{tf3})-(\ref{tot3}), are summarized in Tab.~\ref{tab1}.

\begin{center}
\begin{table}[H]
    \caption{The results in the low-temperature short-distance region $L\ll\lambda\ll \beta$.}
\scalebox{1.1}{
\begin{threeparttable}
\begin{tabular}{|c|c|c|c|}
\hline
$kl$    & $(\delta E)^{kl}_{tf}$ & $(\delta E)^{kl}_{rr}$ &  $(\delta E)^{kl}_{tot}$  \\ \hline
\multirow{2}{*}{$xx$ or $yy$}   & $\frac{\omega_0^5\ln(2\omega_0 L)}{120\pi^2}-\frac{\omega^7_0L^2\ln{(2\omega_0L)}}{140\pi^2}-\frac{(420\gamma-739)\omega^7_0L^2}{58800\pi^2}$ & \multirow{2}{*}{$\frac{9}{4096\pi L^5}-\frac{\omega_0^2}{2048\pi L^3}$} & \multirow{2}{*}{$\frac{9}{4096\pi L^5}$} \\
 & $-\frac{4\pi^6L^2T^8}{525\omega_0}$ &  &   \\ \hline
\multirow{2}{*}{$zz$}    & $-\frac{\omega_0}{64\pi^2L^4}-\frac{\omega_0^3}{96\pi^2L^2}+\frac{\omega_0^5\ln(2\omega_0 L)}{120\pi^2}-\frac{\omega^7_0L^2\ln{(2\omega_0L)}}{420\pi^2}$ & \multirow{2}{*}{$\frac{3}{512\pi L^5}+\frac{\omega_0^2}{256\pi L^3}$}                                                 & \multirow{2}{*}{$\frac{3}{512\pi L^5}$} \\
 & $-\frac{(210\gamma-457)\omega^7_0L^2}{88200\pi^2}-\frac{4\pi^6L^2T^8}{1575\omega_0}$ &  & \\ \hline
\multirow{2}{*}{$xy$}    & $\frac{\omega_0}{256\pi^2L^4}+\frac{\omega_0^3}{384\pi^2L^2}+\frac{\omega_0^5\ln(2\omega_0 L)}{160\pi^2}-\frac{11\omega^7_0L^2\ln{(2\omega_0L)}}{1680\pi^2}$ & \multirow{2}{*}{$\frac{3}{4096\pi L^5}-\frac{3\omega_0^2}{2048\pi L^3}$} & \multirow{2}{*}{$\frac{3}{4096\pi L^5}$} \\
 & $-\frac{(2310\gamma-3977)\omega^7_0L^2}{352800\pi^2}-\frac{11\pi^6L^2T^8}{1575\omega_0}$ & &  \\ \hline
\multirow{2}{*}{$xz$ or $yz$}  & $-\frac{\omega_0}{256\pi^2L^4}+\frac{\omega_0^3}{384\pi^2L^2}-\frac{\omega_0^5\ln(2\omega_0 L)}{160\pi^2}+\frac{\omega^7_0L^2\ln{(2\omega_0L)}}{336\pi^2}$ & \multirow{2}{*}{$\frac{3}{1024\pi L^5}-\frac{\omega_0^4}{512\pi L}$} & \multirow{2}{*}{$\frac{3}{1024\pi L^5}$} \\
  & $\frac{(42\gamma-83)\omega^7_0L^2}{14112\pi^2}+\frac{\pi^6L^2T^8}{315\omega_0}$ & &  \\
\hline
\end{tabular}
\end{threeparttable}}
\label{tab1}
\end{table}
\end{center}

From the second column of Tab.~\ref{tab1}, it is apparent that the first terms are leading and independent of the temperature $T$, thus they dominate the tf-contributions to the interaction, and the temperature-dependent thermal corrections are highly suppressed due to their appearance only at very high orders.
Examining the corresponding forces via the negative derivatives with respect to the object-to-boundary distance $L$, we find that a repulsive gravitational CP force arises exclusively when the object is parallel-planar polarizable (see the fourth row).
In contrast, for the remaining polarization configurations (including the axial and vertical-planar polarizations), the thermal-fluctuation-induced forces are consistently attractive.
Notably, the tf-contribution when the object is parallel-axial polarizable leads to an interaction force with the behavior of $L^{-1}$, which indicates a much slower decay as compared to the $L^{-5}$ scaling observed in the other three polarization cases.

By comparison, the rr-contribution as given by Eq.~(\ref{rr3}), is strictly temperature-independent.
In the limit $L\ll\lambda$, the leading terms in the third column of Tab.~\ref{tab1} consistently yield a repulsive interaction force with the characteristic $L^{-6}$ dependence.
Moreover, since $|(\delta E)^{kl}_{tf}|\ll|(\delta E)^{kl}_{rr}|$ holds for all polarization configurations, the total gravitational CP potentials in the present region $L\ll\lambda\ll \beta$ are effectively governed by the rr-contributions.
Consequently, as shown in the last column of Tab.~\ref{tab1}, the qualitative behavior of the total interaction potentials are largely insensitive to various polarizations of the object.
Nevertheless, quantitative differences persist as the interaction strength is relatively enhanced when the object is polarizable along the axis perpendicular to the boundary.
This indicate that, although the dominate scaling behavior is universal, polarization anisotropy still plays a non-negligible role in determining the magnitude of the gravitational CP interaction in the present region.

\subsubsection{Intermediate-distance region $\lambda\ll L\ll\beta$}

When the object-to-boundary distance $L$ is much larger than the characteristic transition wavelength of the object $\lambda$, the rr-contributions to the gravitational CP potential $(\delta E)_{rr}^{kl}$ are accurately depicted by Eq.~(\ref{rr3}).
It can be clearly seen that, in all polarization configurations, these rr-contributions generally exhibit an oscillatory dependence on the distance $L$.
As a direct consequence of this oscillatory behavior, the corresponding gravitational CP force may convert between attraction and repulsion, and can even vanish at specific distances.

\begin{center}
\begin{table}[H]
    \caption{The results in the low-temperature intermediate-distance region $\lambda\ll L\ll\beta$.}
\scalebox{1.1}{
\begin{threeparttable}
\begin{tabular}{|c|c|c|c|}
\hline
$kl$    & $(\delta E)^{kl}_{tf}$ & $(\delta E)^{kl}_{rr}$ &  $(\delta E)^{kl}_{tot}$  \\ \hline
$xx$ or $yy$   & $\frac{\mathcal{M}_{xxxx}}{1024\pi L^5}+\frac{1}{64\pi^2\omega_0L^6}-\frac{4\pi^6L^2T^8}{525\omega_0}$ & $-\frac{\mathcal{M}_{xxxx}}{1024\pi L^5}$  &   $\frac{1}{64\pi^2\omega_0L^6}-\frac{4\pi^6L^2T^8}{525\omega_0}$ \\  \hline
$zz$    & $\frac{\mathcal{M}_{zzzz}}{1024\pi L^5}+\frac{1}{64\pi^2\omega_0L^6}-\frac{4\pi^6L^2T^8}{1575\omega_0}$ & $-\frac{\mathcal{M}_{zzzz}}{1024\pi L^5}$  & $\frac{1}{64\pi^2\omega_0L^6}-\frac{4\pi^6L^2T^8}{1575\omega_0}$\\  \hline
$xy$    & $\frac{\mathcal{M}_{xyxy}}{1024\pi L^5}+\frac{1}{256\pi^2\omega_0L^6}-\frac{11\pi^6L^2T^8}{1575\omega_0}$ & $-\frac{\mathcal{M}_{xyxy}}{1024\pi L^5}$  & $\frac{1}{256\pi^2\omega_0L^6}-\frac{11\pi^6L^2T^8}{1575\omega_0}$ \\ \hline
$xz$ or $yz$  & $\frac{\mathcal{M}_{xzxz}}{1024\pi L^5}+\frac{1}{256\pi^2\omega_0L^6}+\frac{\pi^6L^2T^8}{315\omega_0}$ & $-\frac{\mathcal{M}_{xzxz}}{1024\pi L^5}$ & $\frac{1}{256\pi^2\omega_0L^6}+\frac{\pi^6L^2T^8}{315\omega_0}$ \\ \hline
\end{tabular}
\begin{tablenotes}
        \footnotesize
        \item[1] The function $\mathcal{M}_{klkl}$ oscillates as the object-to-boundary distance $L$ with the definition $\mathcal{M}_{klkl}\equiv f_{klkl}(\omega_0L)\sin{(2\omega_0 L)}-g_{klkl}(\omega_0L)\cos{(2\omega_0 L)}$ hereafter.
\end{tablenotes}
\end{threeparttable}}
\label{tab2}
\end{table}
\end{center}

The rr-contributions, as well as the approximate tf-contributions $(\delta E)_{tf}^{kl}$, the total interaction potentials $(\delta E)_{tot}^{kl}$ in the intermediate-distance region $\lambda\ll L\ll\beta$ are exhibited in Tab.~\ref{tab2}.
As we can see from the second column in Tab.~\ref{tab2}, the tf-contributions consist of the oscillatory terms, $\frac{\mathcal{M}_{klkl}}{1024\pi L^5}$\;, together with the  monotonic temperature-independent and -dependent terms both significantly smaller than the amplitudes of the oscillatory terms.
As a result, the tf-contributions alone can lead to an either attractive or repulsive and even zero interaction force, depending sensitively on exact values of the distance $L$.
An important feature emerges that the oscillatory terms in the tf-contributions are exactly opposite in sign to those appearing in the rr-contributions, and thus either contribution may dominate in different parameter ranges.

Interestingly, owing to the perfect cancellation of the oscillatory terms in both contributions, the total interaction potentials are determined entirely by the residual non-oscillatory terms from the tf-contributions.
As displayed in the last column of Tab.~\ref{tab2}, the temperature only introduces the minor thermal corrections, and the resulting total potentials in all polarization configurations correspond to a repulsive interaction force that scales as $L^{-7}$.
This distance dependence is one-order higher than that of $L^{-6}$ found in the short-distance region $L\ll\lambda\ll\beta$, and can be attributed to the onset of the retardation effect as the object-to-boundary distance increases beyond the transition wavelength of the object.
It is also found that when the temperature is extremely low and the object-to-boundary distance is relatively larger, the distinctions between different polarization orientations become less pronounced.
In particular, either the parallel-planar(axial) or the vertical-planar(axial) polarization yields the identical leading-order interaction potential, rendering the two mass configurations effectively distinguishable at this level of approximation.
However, to the subleading order, the interaction potential is found to be slightly larger when the mass distribution of the object is vertical to the boundary plane.

\subsubsection{Long-distance region $\lambda\ll\beta\ll L$}

In the long-distance region where the condition $L\gg\lambda$ still holds, the rr-contributions to the gravitational CP potentials $(\delta E)^{kl}_{rr}$ remain completely identical to those in the intermediate-distance region $\lambda\ll L\ll\beta$ and are accurately described by Eq.~(\ref{rr3}), showing a characteristic oscillatory dependence on the distance $L$.

The tf-contributions $(\delta E)^{kl}_{tf}$ in the present region, similar to the case of $\lambda\ll L\ll\beta$, also include both oscillatory and nonoscillatory terms, and the amplitudes of the oscillatory terms are much greater, as can be seen from the second column of Tab.~\ref{tab3}.
When the tf-contributions are combined with the rr-contributions (see the third column of Tab.~\ref{tab3}), the oscillatory terms in them exactly cancel each other, leading to the total interaction potentials governed by the leading-order monotonic terms in the tf-contributions.
Despite these similarities with the region $\lambda\ll L\ll\beta$, the total interaction potentials in the present region remarkably differ in that they exhibit an explicit temperature dependence proportional to $T$, as shown in the last column of Tab.~\ref{tab3}.
By comparing with the total interaction forces in the first two low-temperature regions, $L\ll\lambda\ll\beta$ and $\lambda\ll L\ll\beta$, which scale as $L^{-6}$ and $L^{-7}$, respectively, it becomes evident that when the object-to-boundary distance $L$ greatly exceeds the thermal wavelength $\beta$, the temperature-induced thermal effects become significant and effectively mask the retardation effects.
As a consequence, the scaling of $L^{-6}$ originally characteristic of the short-distance region, reemerges in this long-distance region accompanied by a linear dependence of the temperature.

\begin{center}
\begin{table}[H]
    \caption{The results in the low-temperature long-distance region $\lambda\ll\beta\ll L$.}
\scalebox{1.1}{
\begin{threeparttable}
\begin{tabular}{|c|c|c|c|}
\hline
$kl$    & $(\delta E)^{kl}_{tf}$ & $(\delta E)^{kl}_{rr}$ &  $(\delta E)^{kl}_{tot}$  \\ \hline
$xx$ or $yy$   & $\frac{\mathcal{M}_{xxxx}}{1024\pi L^5}+\frac{9T}{2048\pi\omega_0L^5}-\frac{9}{128\pi^2\omega_0^3L^8}$ & $-\frac{\mathcal{M}_{xxxx}}{1024\pi L^5}$  &   $\frac{9T}{2048\pi\omega_0L^5}$ \\  \hline
$zz$ & $\frac{\mathcal{M}_{zzzz}}{1024\pi L^5}+\frac{3T}{256\pi\omega_0L^5}-\frac{3}{128\pi^2\omega_0^3L^8}$ & $-\frac{\mathcal{M}_{zzzz}}{128\pi L^5}$  & $\frac{3T}{256\pi\omega_0L^5}$ \\  \hline
$xy$ & $\frac{\mathcal{M}_{xyxy}}{1024\pi L^5}+\frac{3T}{2048\pi\omega_0L^5}-\frac{33}{512\pi^2\omega_0^3L^8}$ & $-\frac{\mathcal{M}_{xyxy}}{1024\pi L^5}$  & $\frac{3T}{2048\pi\omega_0L^5}$\\ \hline
$xz$ or $yz$  & $\frac{\mathcal{M}_{xzxz}}{1024\pi L^5}+\frac{3T}{512\pi\omega_0L^5}-\frac{15}{512\pi^2\omega_0^3L^8}$ & $-\frac{\mathcal{M}_{xzxz}}{256\pi L^5}$ & $\frac{3T}{512\pi\omega_0L^5}$ \\ \hline
\end{tabular}
\end{threeparttable}}
\label{tab3}
\end{table}
\end{center}

\subsection{High-temperature regime}

In the high-temperature regime, where the thermal wavelength $\beta$ is much shorter than the characteristic transition wavelength of the object $\lambda$, the gravitational CP interaction exhibits qualitatively distinct behaviors among the regions $L\ll\beta\ll \lambda$, $\beta\ll L\ll\lambda$ and $\beta\ll \lambda\ll L$, which reflects increasingly significant role of the temperature.

\subsubsection{Short-distance region $L\ll\beta\ll \lambda$}

Tab.~\ref{tab4} presents a comprehensive summary, based on Eqs.~(\ref{tf3})-(\ref{tot3}), of the approximate results for the tf-contribution $(\delta E)^{kl}_{tf}$, the rr-contribution $(\delta E)^{kl}_{rr}$ and the total gravitational CP potential $(\delta E)^{kl}_{tot}$ in the high-temperature short-distance region $L\ll\beta\ll \lambda$.

\begin{center}
\begin{table}[H]
    \caption{The results in the high-temperature short-distance region $L\ll \beta\ll\lambda$.}
\scalebox{1.1}{
\begin{threeparttable}
\begin{tabular}{|c|c|c|c|}
\hline
$kl$    & $(\delta E)^{kl}_{tf}$ & $(\delta E)^{kl}_{rr}$ &  $(\delta E)^{kl}_{tot}$  \\ \hline
$xx$ or $yy$   & $\frac{9T}{2048\pi\omega_0 L^5}-\frac{9}{2048\pi L^5}$ & $\frac{9}{4096\pi L^5}-\frac{\omega_0^2}{2048\pi L^3}$  & $\frac{9T}{2048\pi\omega_0 L^5}$ \\ \hline
$zz$    & $-\frac{\omega_0}{64\pi^2L^4}-\frac{\omega^3_0}{96\pi^2L^2}+\frac{\omega^5_0\ln{(2\omega_0 L)}}{120\pi^2}+\frac{4\omega_0L^2T^6}{35\pi^2}$ & $\frac{3}{512\pi L^5}+\frac{\omega_0^2}{256\pi L^3}$                                                 & $\frac{3}{512\pi L^5}$ \\ \cline{1-4}
$xy$    & $\frac{\omega_0}{256\pi^2L^4}+\frac{\omega^3_0}{384\pi^2L^2}-\frac{3\omega^3_0T}{256\pi L}+\frac{11\omega_0L^2T^6}{35\pi^2}$ & $\frac{3}{4096\pi L^5}-\frac{3\omega_0^2}{2048\pi L^3}$ & $\frac{3}{4096\pi L^5}$ \\ \cline{1-4}
$xz$ or $yz$  & $-\frac{\omega_0}{256\pi^2L^4}+\frac{\omega^3_0}{384\pi^2L^2}-\frac{3\omega^3_0T}{256\pi L}-\frac{\omega_0L^2T^6}{7\pi^2}$ & $\frac{3}{1024\pi L^5}-\frac{\omega_0^4}{512\pi L}$ & $\frac{3}{1024\pi L^5}$ \\
\hline
\end{tabular}
\end{threeparttable}}
\label{tab4}
\end{table}
\end{center}

As a consequence of the temperature-independence, the approximate results of the rr-contribution in the present high-temperature region $L\ll\beta\ll \lambda$ coincide with those obtained in the low-temperature short-distance region $L\ll \lambda\ll\beta$, as displayed in the third column of Tab.~\ref{tab4}.
By comparing the tf-contributions in between the two regions (the second columns in Tabs.~\ref{tab1} and~\ref{tab4}), we observe that the temperature plays a more pronounced role in the region $L\ll\beta\ll \lambda$ than in $L\ll \lambda\ll\beta$, as the temperature-dependent terms appear at the lower orders in the high-temperature expansion.
In the specific case where the object is polarizable along the axis parallel to the boundary, the tf-contribution dominated by the thermal modifications exceeds the rr-contribution.
As a result, the total gravitational CP potential exhibits a remarkably strong dependence on the temperature, making the thermal effects a central feature of the interaction in this polarization configuration.
In contrast, for the remaining three polarization configurations, namely when the object is polarizable along the axis perpendicular to the boundary or within a plane, the leading-order total interaction potentials, in comparison with the counterparts in the region $L\ll \lambda\ll\beta$, remain effectively unaltered and temperature-independent.
This consistency is attributed to the dominance of the rr-contributions, outweighing the tf-contributions in which the associated thermal corrections are comparatively negligible in these cases.
Overall, these observations make it clear that the influence led by different orientations of the mass quadrupole moment becomes increasingly prominent.
However, the repulsive property and the characteristic $L^{-6}$-dependence of the total gravitational CP force are universally valid among all the polarization configurations.

\subsubsection{Intermediate-distance region $\beta\ll L\ll\lambda$}

In accordance with the results summarized in Tab.~\ref{tab5}, both the tf-contributions to and the total gravitational CP potentials in the high-temperature intermediate-distance region $\beta\ll L\ll\lambda$ can be drastically influenced by temperature, and compared with the behaviors observed in the regions $L\ll\lambda\ll \beta$ and $L\ll \beta\ll\lambda$, a variety of novel features occur.

\begin{center}
\begin{table}[H]
    \caption{The results in the high-temperature intermediate-distance region $ \beta\ll L\ll\lambda$.}
\scalebox{1.1}{
\begin{threeparttable}
\begin{tabular}{|c|c|c|c|c|c|c|}
\hline
$kl$    & $(\delta E)^{kl}_{tf}$ & $(\delta E)^{kl}_{rr}$ &  $(\delta E)^{kl}_{tot}$  \\ \hline
$xx$ or $yy$ & $\frac{\omega_0 T}{1024\pi L^3}-\frac{3\omega_0^3T}{1024\pi L}-\frac{\omega_0^5\ln{(2\omega_0L)}}{120\pi^2}$ & $\frac{9}{4096\pi L^5}-\frac{\omega_0^2}{2048\pi L^3}$  &   $\frac{9}{4096\pi L^5}+\frac{\omega_0 T}{1024\pi L^3}$ \\  \hline
$zz$ & $-\frac{\omega_0 T}{128\pi L^3}-\frac{\omega_0^3T}{128\pi L}+\frac{\omega_0}{64\pi^2 L^4}$ & $\frac{3}{512\pi L^5}+\frac{\omega_0^2}{256\pi L^3}$  & $\frac{3}{512\pi L^5}-\frac{\omega_0 T}{128\pi L^3}$ \\  \hline
$xy$ & $\frac{3\omega_0 T}{1024\pi L^3}-\frac{\omega_0^3T}{1024\pi L}-\frac{\omega_0}{256\pi^2 L^4}$ & $\frac{3}{4096\pi L^5}-\frac{3\omega_0^2}{2048\pi L^3}$  & $\frac{3}{4096\pi L^5}+\frac{3\omega_0 T}{1024\pi L^3}$ \\ \hline
$xz$ or $yz$  & $\frac{\omega_0^3T}{256\pi L }+\frac{\omega_0}{256\pi^2 L^4}$ & $\frac{3}{1024\pi L^5}-\frac{\omega_0^4}{512\pi L}$ & $\frac{3}{1024\pi L^5}+\frac{\omega_0^3T}{256\pi L }$ \\
\hline
\end{tabular}
\end{threeparttable}}
\label{tab5}
\end{table}
\end{center}

First of all, the effects on the tf-contributions induced solely by zero-point fluctuations of gravitational fields, when set against the temperature-induced thermal corrections, are comparatively weak regardless of the polarizations of the object, as the dominating terms are all explicitly temperature-dependent.
Second, a comparison between the tf-contributions in the present high-temperature region $\beta\ll L\ll\lambda$ and those in the low-temperature region $L\ll\lambda\ll \beta$, i.e. the second columns in Tabs.~\ref{tab1} and~\ref{tab5}, reveals that the first terms in the region $L\ll\lambda\ll \beta$, which are leading and temperature-independent, undergo a clear sign reverse upon entering the region $\beta\ll L\ll\lambda$.
This sign change suggests that the temperature-independent thermal corrections can also be generated at the extremely high temperature.
By adding up the tf- and rr-contributions, we further find that the detailed behaviors of the total gravitational CP interactions, as shown in the last column of Tab.~\ref{tab5}, depend not only on the polarizations of the object but also sensitively on the concrete ranges of the distance $L$.
Specifically, in the region $\beta\ll L\ll\sqrt[4]{\beta\lambda^3}$ and when the object is vertical-planar polarizable, the total gravitational CP potential is primarily contributed by the radiation reaction of the object and corresponds to a repulsive force proportional to $T^0L^{-6}$.
The same conclusion also holds for the other three polarization configurations within the range $\beta\ll L\ll\sqrt{\beta\lambda}$.
Outside these cases, however, the tf-contributions dominate the interaction, and in contrast to a universal repulsive force behaving as $L^{-6}$, the qualitatively new behaviors are present due to the temperature-induced effects.
On the one hand, the novel distance dependence appears, $\sim L^{-2}$ in the situation with $\sqrt[4]{\beta\lambda^3}\ll L\ll\lambda$ and the object vertical-planar polarizable, and $\sim L^{-4}$ in the other polarization situations with $\sqrt{\beta\lambda}\ll L\ll\lambda$.
These results demonstrate that the thermal effects can fundamentally alter the distance scaling of the interaction.
On the other hand, especially when the object is vertical-axial polarizable, the gravitational CP force becomes uniquely attractive in the subregion $\sqrt{\beta\lambda}\ll L\ll\lambda$.

\subsubsection{Long-distance region $\beta\ll\lambda\ll L$}

In Tab.~\ref{tab6}, we present the approximated results for the gravitational CP potential in the long-distance region $\beta\ll \lambda\ll L$, which also corresponds to an extremely high-temperature limit.
Similar to the high-temperature intermediate-distance region $\beta\ll L\ll\lambda$, the temperature-induced effects manifest themselves in two aspects.
On the one hand, as displayed in the second column of Tab.~\ref{tab6}, the temperature-dependent modifications due to their relatively larger amplitudes, play a dominant role in the tf-contributions and are generally larger than the corresponding rr-contributions.
And the total interaction potentials mainly attributed to the tf-contribution, which decay monotonically in the region $\beta\ll L\ll\lambda$, however exhibit an oscillatory dependence on the distance $L$ now.
As a result, the corresponding forces can be either attractive or repulsive and, notably, even vanishing at special distances.
On the other hand, by comparing the temperature-independent oscillatory terms of the tf-contributions in the low-temperature regions $\lambda\ll L \ll\beta$ and $\lambda\ll\beta\ll L$ with those in the present high-temperature region $\beta\ll \lambda\ll L$ (refer to Tabs.~\ref{tab2},~\ref{tab3} and~\ref{tab6}), a noticeable difference in sign can be observed, which is totally induced by the temperature-independent modifications generated at the very high temperature.
Consequently, the exact cancellation of the temperature-independent oscillatory terms in the tf- and rr-contributions, which occurs in the low-temperature intermediate- and long-distance regions, is no longer possible in the present high-temperature region.

\begin{center}
\begin{table}[H]
    \caption{The results in the high-temperature long-distance region $\beta\ll\lambda\ll L$.}
\scalebox{1.1}{
\begin{threeparttable}
\begin{tabular}{|c|c|c|c|}
\hline
$kl$    & $(\delta E)^{kl}_{tf}$ & $(\delta E)^{kl}_{rr}$ &  $(\delta E)^{kl}_{tot}$  \\ \hline
$xx$ or $yy$   & $\frac{\frac{2T}{\omega_0}\left(\mathcal{M}_{xxxx}+\frac{9}{4}\right)-\mathcal{M}_{xxxx}}{1024\pi L^5}-\frac{1}{64\pi^2\omega_0L^6}$ & $-\frac{\mathcal{M}_{xxxx}}{1024\pi L^5}$  &   $\frac{T\left(\mathcal{M}_{xxxx}+\frac{9}{4}\right)}{512\pi\omega_0 L^5}$ \\  \hline
$zz$    & $\frac{\frac{2T}{\omega_0}\left(\mathcal{M}_{zzzz}+\frac{3}{2}\right)-\mathcal{M}_{zzzz}}{128\pi L^5}-\frac{1}{64\pi^2\omega_0L^6}$ & $-\frac{\mathcal{M}_{zzzz}}{128\pi L^5}$  & $\frac{T\left(\mathcal{M}_{zzzz}+\frac{3}{2}\right)}{64\pi \omega_0L^5}$ \\  \hline
$xy$    & $\frac{\frac{2T}{\omega_0}\left(\mathcal{M}_{xyxy}+\frac{3}{2}\right)-\mathcal{M}_{xyxy}}{1024\pi L^5}-\frac{3}{256\pi^2\omega_0L^6}$ & $-\frac{\mathcal{M}_{xyxy}}{1024\pi L^5}$  & $\frac{T\left(\mathcal{M}_{xyxy}+\frac{3}{2}\right)}{512\pi\omega_0 L^5}$\\ \hline
$xz$ or $yz$  & $\frac{\frac{2T}{\omega_0}\left(\mathcal{M}_{xzxz}+\frac{3}{2}\right)-\mathcal{M}_{xzxz}}{256\pi L^5}-\frac{3}{256\pi^2\omega_0L^6}$ & $-\frac{\mathcal{M}_{xzxz}}{256\pi L^5}$ & $\frac{T\left(\mathcal{M}_{xzxz}+\frac{3}{2}\right)}{128\pi L^5}$ \\ \hline
\end{tabular}
\end{threeparttable}}
\label{tab6}
\end{table}
\end{center}

\section{Summary}

We have systematically investigated the gravitational Casimir-Polder (CP) interaction between a gravitationally polarizable two-level object and an infinite gravitational Dirichlet boundary in a thermal bath at the temperature $T$.
With the use of the DDC formalism, we separated the gravitational CP interaction into the contributions of gravitational thermal fluctuations (tf) and the radiation reaction of the object (rr).
The results indicate that the rr-contributions are independent of the temperature and coincide with the counterparts in vacuum case, which monotonically decay as $L^{-5}$ when $L\ll\lambda$, with $L$ the object-to-boundary distance and $\lambda$ the characteristic transition wavelength of the object.
When $L\gg\lambda$, however, the rr-contributions exhibit the oscillatory behaviors, giving rise to either attractive or repulsive and even vanishing interaction forces, in sharp contrast to the universally repulsive force at extremely short distances.
The strengths of the interaction potential are also found to depend weakly on the orientations of mass quadrupole moment of the object.

The temperature affects both the tf-contributions and the total interaction potentials, with their behaviors at high temperatures determined by a nontrivial interplay between the thermal and the polarization effects.
In the low-temperature regime $\lambda\ll\beta$, where $\beta$ is the thermal wavelength of gravitons, three distance regions can be distinguished, i.e.,  $L\ll\lambda\ll\beta$, $\lambda\ll L\ll\beta$ and $\lambda\ll\beta\ll L$.
In the first two regions, i.e. when $L\ll\beta$, the temperature-induced thermal corrections are negligible.
In the short-distance region $L\ll\lambda\ll\beta$, the tf-contributions are subdominant to the rr-contributions scaling as $L^{-5}$; in the intermediate-distance region $\lambda\ll L\ll\beta$, either contribution may dominate owing to their respective oscillatory behaviors, while interestingly the total potentials scale as $L^{-6}$ since the oscillatory terms in the two contributions cancel exactly.
Although this cancellation also occurs in the third long-distance region with $L$ exceeding $\beta$, the thermal corrections then become dominant and lead to a scaling of $TL^{-5}$ for the interaction potentials.

The high-temperature regime exhibits qualitatively different behaviors, with the polarization configurations of the object affecting not only the magnitude but also the distance dependence of the interaction.
The relevant high-temperature regions are  $L\ll\beta\ll\lambda$, $\beta\ll L\ll\lambda$ and $\beta\ll\lambda\ll L$.
When $L\ll\beta\ll\lambda$, the tf-contribution controlled by the thermal effect overweighs the rr-contribution when the object is polarizable along the axis parallel to the boundary plane, resulting in a total potential proportional to $TL^{-5}$.
However, in other polarization configurations, including those with the mass distributions along the axis vertical to the boundary or within a plane, the temperature-independent rr-contributions are more significant.

Novel distance scalings arise in the high-temperature intermediate-distance region $\beta\ll L\ll\lambda$.
For a vertical-planar polarizable object, the total potentials behave as $TL^{-1}$ in the subregion $\sqrt[4]{\beta\lambda^3}\ll L\ll\lambda$, whereas they scale as $TL^{-3}$ in the other three polarization situations when $\sqrt{\beta\lambda}\ll L\ll\lambda$.
Notably, a uniquely attractive interaction force is found for the vertical-axial polarization.
In the final long-distance region $\beta\ll \lambda\ll L$, the thermal corrections associated with the tf-contributions uniquely produce the oscillatory-varying interaction potentials, sharply distinctive from the universally monotonic behaviors found in the other regions.
Meanwhile, another remarkable effect arising at the high temperatures is the sign reversal of the temperature-independent oscillatory terms in the tf-contributions, which consequently prevents their cancellation with the corresponding terms in the rr-contributions.

\begin{acknowledgments}

This work was supported by the NSFC under Grants No. 12047551 and No. 12105061.

\end{acknowledgments}

\end{document}